# Germagraphene as promising anode material for Lithium-ion batteries predicted from first-principles calculations

Junping Hu,[a,b] Chuying Ouyang,[c] Shengyuan A. Yang,[*a,d] Hui Ying Yang[*a]

Finding electrode materials with high capacity is a key challenge for developing Lithium-ion batteries (LIBs). Graphene was once expected to be a promising candidate, but it turns out to be too inert to interact with Li. Here, by using the first-principles calculations, we predict that Germanium doped graphene, termed as Germagraphene which has been achieved in recent experiment, is a promising LIB anode material. We find that at the optimal Ge concentration, which corresponds to the chemical formula $C_{17}$Ge, the specific capacity for Germagraphene can be as high as 1734 mAh/g, over four times larger than graphite. We show that the material is conductive before and after Li adsorption. We also investigate the diffusion process of Li on Germagraphene, and find that the diffusion barrier is low (∼0.151 eV), implying fast Li diffusion. The calculated average intercalation potential is very low (∼0.03 V), which is beneficial for increasing the working voltage for the full-cells. In addition, during the process of Li intercalation, the lattice change for the material is quite small (∼0.48%), implying good cycle performance. These results suggest that Germagraphene could be a promising high-capacity anode material for LIBs.

## 1. Introduction

From smart phones to laptops and to electric vehicles, lithium-ion batteries (LIBs) are powering a wide range of electronics. However, as consumer technology becomes more powerful, LIB technology has always been struggling to keep pace. Researchers have made significant effort to achieve cost-effective, long-cycle-life, and especially high-capacity energy storage systems. For example, a new strategy is to use a sulfur positive electrode and a lithium-metal negative electrode to increase the capacity [1,2]. At the same time, with the rapid development of two-dimensional (2D) materials, there is increasing interest in exploring 2D materials as possible electrode materials[3-10], owing to their unique 2D structures, possibly high electric conductivity, and most importantly, the large specific surface area. It has been demonstrated that 2D materials typically can store large amount of Li or other metal ions, to achieve very high storage capacities. In addition, 2D materials may also enjoy the advantages of rapid ion migration and relatively small volume change during ion intercalation and de-intercalation processes. These two factors are critical for rate performance and for maintaining the structural integrity.

Among the many 2D materials studied to date, graphene is undoubtedly the most prominent one. It has high mobility, excellent flexibility, and good structural and electrochemical stability, which is comparable to or better than the commercial anode materials, but a severe disadvantage of graphene is that it is too inert to interact with Li, so the storage capacity is low. In order to enhance the interaction between graphene and Li, researchers have applied various modification methods on graphene, such as introducing defects or functional groups[11-15], which has indeed improved the Li storage capacity to some extent; however, the conductivity, the ion diffusion rate, the first cycle Coulomb efficiency, and the structural stability during the cycle are often degraded in the meantime. These shortcomings severely limit the wide application of graphene-based nanomaterials for LIBs. Another approach is via the same-group doping, i.e., to replace a percentage of carbon atoms in graphene by elements from the same group. In a recent work, Wang *et al.* proposed a new 2D material called Siligraphene which can be viewed as graphene with Si doping. They found that the interaction between Li ions and the host is greatly enhanced, resulting in a very high Li storage capacity (theoretical value up to ∼1520 mAh/g) [16]. The proposed material not only avoids the volume expansion problem for bulk silicon, but could also be free from the stability issue with silicene under ambient conditions.

[a.] Research Laboratory for Quantum Materials, Singapore University of Technology and Design, Singapore 487372, Singapore
[b.] School of Science, Nanchang Institute of Technology, Nanchang 330099, China
[c.] Department of Physics, Laboratory of Computational Materials Physics, Jiangxi Normal University, Nanchang 330022, China.
[d.] Center for Quantum Transport and Thermal Energy Science, School of Physics and Technology, Nanjing Normal University, Nanjing 210023, China

Most Recently, Tripathi *et al.* reported that they succeeded in doping Ge into graphene by using the low-energy ion implantation technique [17], and the doping concentration can be controlled in experiment. We note that similar to the bulk silicon, Germanium in the bulk form has also been tested for an electrode material, and high storage capacity has been demonstrated; however, the large volume expansion and the issue of oxidation in air limit its applications [18]. Inspired by Siligraphene, one naturally wonders: Will the Ge implanted graphene, which may be termed as Germagraphene, be a promising 2D electrode material?

Motivated by the experimental breakthrough mentioned above, in this paper, we investigate the performance of Germagraphene as an electrode material for LIBs via first-principles calculations. We find that Germagraphene can achieve a high Li storage capacity, with a maximum value of about 1734 mAh/g for the optimal doping concentration (corresponding to a chemical formula of $C_{17}$Ge). The material remains conductive before and after Li adsorption, and it has low Li diffusion barrier ~0.151 eV, which are desired for a good electrode material. The calculated average intercalation potential is very low (~0.03 V), which would be beneficial for expanding the working voltage window for the full-cells. In addition, during the process of Li intercalation, the lattice change of $C_{17}$Ge is quite small (~0.48%), indicating that the material should have good cycle performance. Thus, our work reveals the great potential of Germagraphene as electrode materials, and it also offers new insight in engineering 2D materials for energy storage applications.

## 2. Computational details

Our first-principles calculations are based on the density functional theory (DFT) using the plane-wave pseudopotentials [19-20] as implemented in the Vienna *ab initio* Simulation Package (VASP) [21-22]. The exchange-correlation effect is modelled in the local density approximation (LDA) with the Ceperly-Alder functional [23] parameterized by Perdew and Zunger [24]. Carbon $2s^2 2p^2$ and Germanium $4s^2 4p^2$ electrons are treated as valence electrons in all the calculations. A cutoff energy of 500 eV is employed for the plane wave expansion. The Brillouin zone is sampled with 5 × 5 × 1 Monkhorst-Pack $k$-point mesh [25] for the structural optimization, and with 7 × 7 × 1 mesh for the electronic structure calculations. The convergence criteria for the total energy and ionic forces are set to be $10^{-5}$ eV and $10^{-3}$ eV/Å, respectively. Phonon spectra calculations are performed using the Phonopy package [26]. In our calculations, periodic images of the monolayer along the plane normal direction are separated by a vacuum layer greater than 18.0 Å, so that the interaction between the images is negligible.

It should be mentioned that we adopt the LDA functional instead of the GGA functional, because LDA usually gives a better description for 2D carbon materials due to an internal double error elimination [27-28]. It has been reported that the LDA Ceperly-Alder functional gave more reasonable results for the interaction between Li and 2D carbon materials than GGA functionals [16,29].

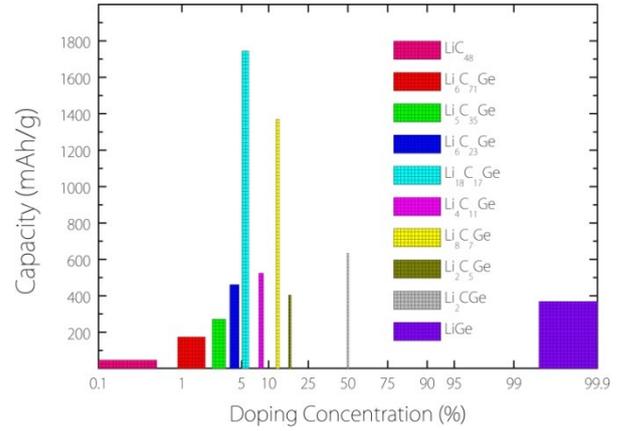

Figure 1. Plot of Li storage capacity under different doping concentrations

The adsorption energy ($E_{ad}$) for Li atom on the Germagraphene monolayer is defined as:

$$E_{ad} = (E_{C_xGe_yLi_z} - E_{C_xGe_y} - zE_{Li})/z \qquad (1)$$

where $E_{Li}$ is the cohesive energy for the Li metal, $E_{C_xGe_yLi_z}$ and $E_{C_xGe_y}$ are the total energies for the Germagraphene $C_xGe_y$ monolayer with and without Li adsorption. The "$z$" here indicates the amount of the adsorbed Li atoms.

The difference in total energies before and after Li intercalation is used to determine the average intercalation potential. The volume and entropy effects are usually negligible during the reaction, hence they are omitted. Then the average intercalation potential for a reaction involving $z$Li$^+$ ions can be approximately calculated from the energy difference:

$$V_{ave} = -(E_{C_xGe_yLi_z} - E_{C_xGe_y} - zE_{Li})/ze \qquad (2)$$

## 3. Results and discussion

### 3.1 Ge concentration and Li storage capacity

From the experimental observation, it has been found that an implanted Ge atom could directly substitute a C atoms in graphene, bonding to three carbon neighbours and forming a buckled out-of-plane structure, which is also predicted by DFT calculations [17]. Based on this kind of buckled out-of-plane configuration, we estimate the Li storage capacity under different Ge doping concentrations for a Germagraphene monolayer. Our results are shown in Fig. 1. Here, we have investigated totally 10 concentrations, namely, 0%, 1.39% (1/72), 2.78% 2.78% (1/36), 4.18% (1/24), 5.56% (1/18), 8.33% (1/12), 12.50% (1/8), 16.67% (1/6), 50% (1/2), 100% (1/1). For modelling the different concentration, two different supercells that contain 6 and 8 carbon atoms in graphene are adopted (see Figure S1 in the Supporting Information). Each model is fully relaxed before and after Li intercalation.

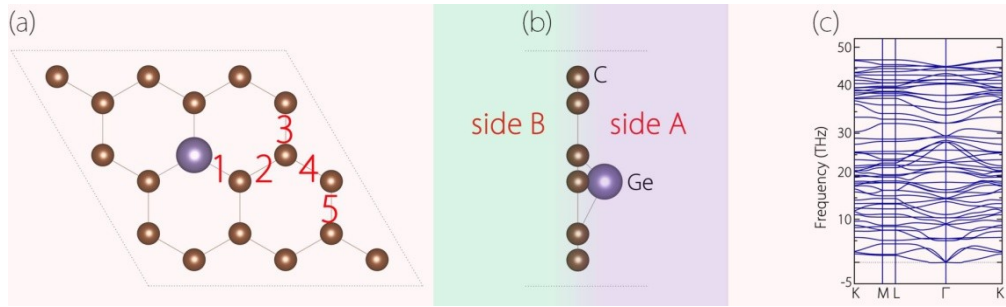

Figure 2. (a) Top view and (b) side view of the optimized $C_{17}Ge$. (c) Calculated phonon spectrum for $C_{17}Ge$.

For each model, the adsorption energy for each adsorption site (see Fig. 3) is calculated by Equation (1). If the adsorption is energetically favourable, then the amount of adsorbed Li is further increased. When the average Li insertion voltage becomes close to 0 V, the number of Li intercalation numbers is taken as the basis for calculating the Li storage capacity. For the concentration of 0%, it actually corresponds to the pristine graphene. For this case, the calculated Li-adsorption ratio 1/48 is even smaller than the Li/C ratio for a completely exhausted graphite anode ($LiC_6$, corresponding to a capacity of 372 mAh/g). This is consistent with the previous finding that the Li-graphene interaction is very weak [16].

The capacity as a function of Ge concentration in Fig. 1 exhibits a quasi-normal distribution. The highest capacity value (>1700 mAh/g) is reached at a doping concentration of 1/18, and the corresponding chemical formula is $C_{17}Ge$. In the following discussion, we shall focus on this optimal case.

### 3.2 Structure of Germagraphene $C_{17}Ge$

Before investigating the energy storage properties of Germagraphene $C_{17}Ge$, we first give a detailed examination of its crystal structure. In the structure, there are totally five different bond lengths, as labelled in Fig. 2(a). The lengths of bond 1 to bond 5 are 1.845 Å, 1.401 Å, 1.448 Å, 1.412 Å, and 1.433 Å, respectively. The average bond length here is longer than the carbon-carbon bond length of 1.420 Å in graphite [30]. As we mentioned before, Ge is located slightly above the C plane. The height of the Ge atom relative to C plane is about 0.72 Å. The buckling breaks the mirror symmetry of the structure. In Fig. 2(b), we label the two sides of the monolayer plane as Side A and Side B.

We have also investigated the dynamical stability of the structure, which can be inferred from its phonon spectrum. Figure 2(c) shows the phonon spectrum for $C_{17}Ge$. Clearly, there is no imaginary frequency mode, which demonstrates that the structure is dynamically stable.

### 3.2 Li adsorption

The conductivity of the electrode material is critical to the rate performance of LIBs. It is preferable that the material remains metallic before and after Li adsorption.

To study the Li adsorption on Germagraphene, we first need to determine the most favourable adsorption site for the Li atom. The possible adsorption sites that we have tested are shown in Fig. 3(a). These sites can be classified into three types [see Fig. 3(a)]: the bridge site between two atoms, the top site on an atom, and the hollow site above a hexagonal ring. Moreover, since the two sides of Germagraphene are asymmetric, for each type of site, we also need to consider the two sides A and B separately.

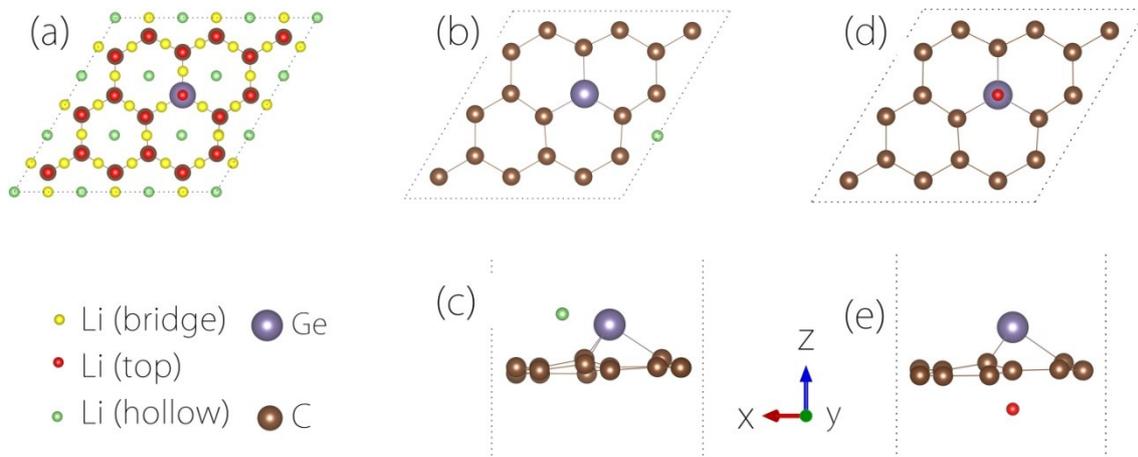

Figure 3. (a) Top view of all adsorption sites on $C_{17}Ge$ that we have considered. (b) Top view and (c) side view of the most stable adsorption configuration with Li atom on Side A. (d) Top view and (e) side view of the most stable adsorption configuration with Li atom on Side B.

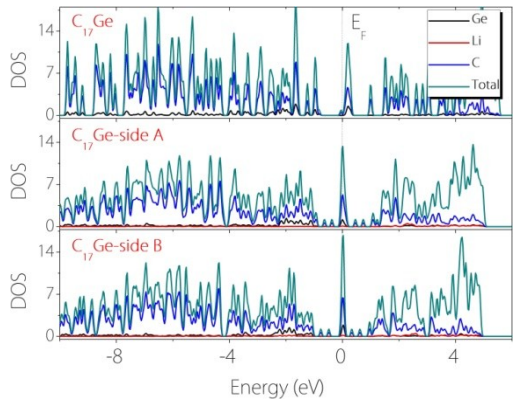

Figure 4. Density of states for (a) pristine $C_{17}Ge$, (b) $LiC_{17}Ge$ (adsorption on Side A), and (c) $LiC_{17}Ge$ (adsorption on Side B).

The preference for Li adsorption at these sites is characterized by the adsorption energy: The lower the adsorption energy, the more stable the adsorption configuration. For adsorption on Side A, the most stable adsorption site is near a hollow site above a $C_6$ ring [see Fig. 3(b) and (c)]. The corresponding adsorption energy is 0.57 eV. For adsorption on Side B, The most stable adsorption site is the top site above Ge [see Fig. 3(d) and (e)], and the corresponding adsorption energy is about 0.94 eV. During the adsorption, there is charge transfer between Li and the Germagraphene layer. For example, for the case in Fig. 3(d,e), we find from the Bader charge analysis that both Li and Ge lose ~0.85e charge and become positively charged; while each C atom obtains ~0.09e and becomes negatively charged.

To check whether the material can remain conductive before and after Li adsorption, we calculate the density of states (DOS) before and after Li adsorption and results are plotted in Fig. 4. One can observe that the Germagraphene $C_{17}Ge$ is intrinsically metallic, and the DOS near the fermi level is dominated by the contribution from C orbitals. Upon Li adsorption, the system is still metallic for the two configurations discussed above. Furthermore, the DOS at the Fermi level is even enhanced by Li adsorption, suggesting that more carriers are generated in the system. As a result, the electronic conduction should become better, which is beneficial for LIBs.

### 3.3 Li diffusion barrier

Next, we turn to study the Li diffusion process on Germagraphene, which is critical for the rate performance of LIBs. In order to reduce the artificial interaction effect between neighbouring images, a 2 × 2 supercell containing 68 C atom and 4 Ge atoms is employed for the calculation. The climbing-image nudged elastic band (NEB) method [31] is used to determine the diffusion energy barrier height, and to seek the optimal diffusion pathways as well as the saddle points. Again, due to the asymmetry between Side A and Side B, the diffusion processes on the two sides are considered separately.

For Side A, we consider three possible migration pathways, according to the movement of Li atom between two neighbouring most stable adsorption sites. The three pathways are marked in the shadowed region in Fig. 5(a). The images for the Li atom along the paths are indicated by red (Path-1), green (Path-2), and blue (Path-3) colours, respectively. It can be clearly seen that the three paths all become curved after NEB calculation, no longer straight as they are set initially. The calculated energy barriers are plotted in Fig. 5(b), which are 0.194, 0.193, and 0.151 eV, respectively. These three values are all smaller than the Li diffusion barriers for pristine graphene (~0.277 eV), graphene with point defect (~0.366 eV to 0.538 eV) [32], as well as graphite (~0.47 eV) [33], promising a fast Li diffusion. For the Side B, since the most stable adsorption site is the top site of Ge, only one migration pathway needs to be considered [see Fig. 5(a)]. The corresponding barrier height is found to be 0.837 eV, as shown in Fig. 5(c). Although this value is higher than Side A and graphene, it is still acceptable for a practical electrode material.

### 3.4 Li storage capacity

The Li storage capacity and the average intercalation potential are the key characteristics for the LIB electrode materials. To study these properties, based on our discussion in the previous section 3.2, we increase the concentration of adsorbed Li atoms on Germagraphene. In estimating the maximum possible capacity for Li adsorption, we still use the 18-atom unit cell in Fig. 3(a), and increase the number of Li atoms adsorbed on both sides of the monolayer. The average intercalation potential can be approximately calculated from the energy difference as in Equation (2) for the intercalation reaction involving $z$ Li ions:

$$C_{17}Ge + zLi^+ + ze^- \leftrightarrow Li_zC_{17}Ge \quad (3)$$

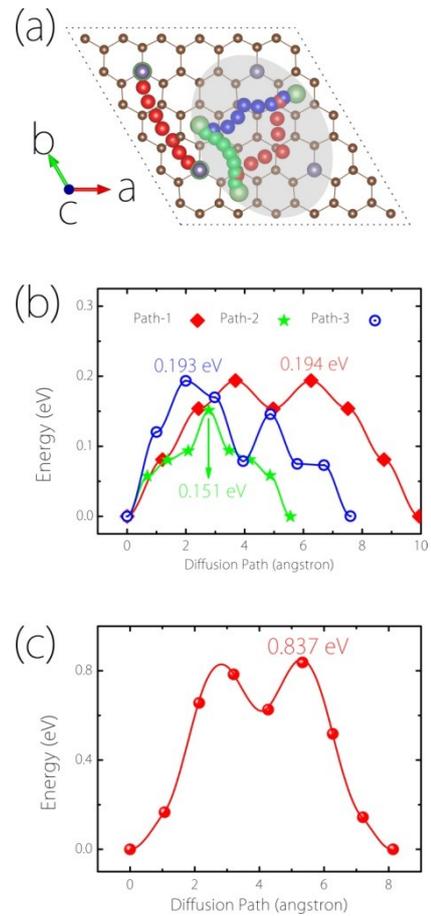

Figure 3. (a) Diffusion pathways. Three pathways in the shaded region are on Side A and the other one is on Side B. (b) Diffusion barriers for the pathways on Side A. (c) Diffusion barrier for the pathway on Side B.

Starting from occupying the most stable adsorption site, the number of adsorbed Li atoms is increased one by one. At each step, we scan all possible sites again to locate the most stable one for the next adsorbed Li atom. We calculate the average Li insertion voltage according to Equation (2), and at the same time, we calculate the sequential adsorption energy, as defined below:

$$E_{sae} = E_{C_{17}GeLi_{m+1}} - E_{C_{17}GeLi_m} - E_{Li} \quad (4)$$

where $E_{Li}$ is the cohesive energy for Li metal, $E_{C_{17}GeLi_{m+1}}$ and $E_{C_{17}GeLi_m}$ are the total energies of Germagrahpene with m+1 and m Li, respectively. The negative value of $E_{sae}$ and the positive value of $V_{ave}$ would indicate that the corresponding Li atom can be intercalated. Since Li ions typically migrate rapidly during the charging and discharging processes, we only consider the five configurations with m being equal to 6, 12, 18, and 20. The calculated values for the average intercalation potential are sequentially -0.19 V, -0.12 V, -0.04 V, and 0.18 V, which indicates that the Germagraphene $C_{17}Ge$ as a LIB anode material can hold up to 18 Li atoms in a unit cell. To further validate this, we calculate the sequential adsorption energies, and the results show that they are all positive values, which proves that $C_{17}Ge$ can indeed accommodate 18 Li atoms per unit cell. The corresponding chemical stoichiometry is $Li_{18}C_{17}Ge$. The theoretical specific capacity is thus 1734 mAh/g, which is ~4.7 times that of graphite (~372 mAh/g for $LiC_6$). Meanwhile, we find that the calculated average intercalation potential is very low (~0.03 V), which is beneficial for expanding the working voltage window for the full-cells.

The capacity for Germagraphene is very high. We can compare this value with several other 2D anode materials proposed so far (see Fig. 6). This value for Germagraphene is 5–7 times that for $MoS_2$/graphene (~335 mAh/g (ref. 19)) and GeS nanosheets [34] (~256 mAh/g (ref. 59)); 3–4 times that for most MXenes, including $Nb_2C$ [35] (~542 mAh/g), $Mo_2C$ [36] (~526 mAh/g) and $Ti_3C_2$ [37,38] (~319 mAh/g); and almost 2 times that of the recently reported silicone [39] (~954 mAh/g) and $V_2C$ [40] (~941 mAh/g). This capacity is only slightly lower than that for the $\beta_{12}$-borophene (~1984 mAh/g) [8] but is even larger than the capacity for $\chi_3$-borophene (~1240 mAh/g) [8] and Siligraphene (~1520 mAh/g (g-$SiC_5$)) [16].

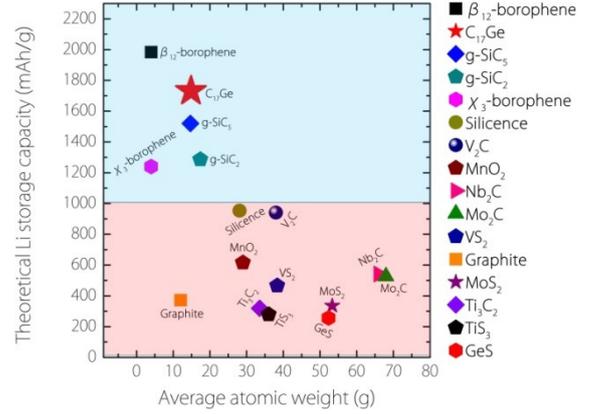

Figure 4. Comparison of the theoretical Li storage capacity values between Germagraphene $C_{17}Ge$ and other 2D anode materials (plus graphite).

The discussion above shows that Germagraphene can be a high capacity anode material for LIBs. The enhanced interaction between Li and Germagraphene increases the charge and discharge voltages, and also allows a large amount of Li to be absorbed. Obviously, the presence of Ge is the reason for the enhanced interaction with Li. To further see this, we perform the Bader charge analysis for the system before and after Li adsorption. The results are listed in Table 1. As can be seen, the Ge atom in $C_{17}Ge$ loses ~0.96e and becomes charged positively, because the electronegativity of C (2.55) is larger than that of Ge (2.01). As the Li intercalation process progresses, the total number of gained electrons for C atoms is always greater than the number of electron loss for the Ge atom, and the Li atoms are always in the state of electron loss, indicating that the presence of Ge is beneficial for the Li adsorption.

## 4. Conclusions

In conclusion, by using first-principles calculations, we demonstrate that Germagraphene can be a promising high-capacity LIB anode material. We find that the capacity at the optimal Ge concentration ($C_{17}Ge$) can reach ~1734 mAh/g, larger than most other 2D materials studied so far. We show that the material has good electric conduction before and after Li adsorption. The Li diffusion barrier can be as low as 0.151 eV, indicating fast rate-performance. The calculated average intercalation potential is very low (~0.03 V), which is beneficial for expanding the working voltage window for the full-cells. Finally, in the process of Li intercalation, we find that the lattice change of $C_{17}Ge$ is quite small (only about 0.48%), indicating the material should have good cycle performance as well. Experimentally, Germagraphene with controlled Ge concentration has been demonstrated by using the low-energy ion implantation, which is expected to be a scalable and precise technique for controllably doping 2D materials [17]. Our work not only reveals a promising candidate material with ultrahigh capacity for LIBs, it also offers a new insight in engineering 2D materials for energy storage applications.

Table 1. Bader charge analysis for $C_{17}Ge$ and its Li intercalated states.

| | Average Charge State | | |
|---|---|---|---|
| | Li | C | Ge |
| $C_{17}Ge$ | - | 4.05(-0.05) | 3.04(+0.96) |
| $LiC_{17}Ge$ | 0.15(+0.85) | 4.09(-0.09) | 3.15(+0.85) |
| $Li_6C_{17}Ge$ | 0.27(+0.73) | 4.27(-0.27) | 3.78(+0.22) |
| $Li_{12}C_{17}Ge$ | 0.29(+0.71) | 4.51(-0.51) | 3.93(+0.07) |
| $Li_{18}C_{17}Ge$ | 0.35(+0.65) | 4.68(-0.68) | 4.18(-0.18) |

## Conflicts of interest

There is no conflict of interest to declare.

## Acknowledgements

The authors thank D. L. Deng for valuable discussions. This work is supported by the National Research Foundation Prime Minister's Office Singapore under NRF-ANR Joint Grant Call (Award No. NRF2015-NRF-ANR000-CEENEMA), Natural Science Foundation of China (Grant No. 11564016), and Singapore Ministry of Education Academic Research Fund Tier 2 (MOE2015-T2-2-144). We acknowledge computational support from the Texas Advanced Computing Center and the National Supercomputing Centre Singapore.

## Notes and references